# The improved ARTEMIS IV
# multichannel solar radio spectrograph of the University of Athens

A. Kontogeorgos[1,2], P. Tsitsipis[1,2], C. Caroubalos[1], X. Moussas[1], P. Preka-Papadema[1],
A. Hilaris[1], V. Petoussis[1,2], J.-L. Bougeret[3], C. E. Alissandrakis[4], G. Dumas[3],

[1] Dep. Of Physics and Dep. Of Informatics, University of Athens, GR_15783, Athens, Greece
kkarou@di.uoa.gr, xmoussas@cc.uoa.gr, ppreka@cc.uoa.gr, ahilaris@cc.uoa.gr,
[2] Dep. Of Electronics, Technological Education Institute of Lamia, GR-35100, Lamia, Greece
akontog@teilam.gr, tsitsipis@teilam.gr, petos@teilam.gr
[3] Observatoire de Paris, LESIA, CNRS UA 264, F-92195 Meudon Cedex, France
superdumas@libertysurf.fr, bougeret@obspm.fr
[4] Dep. Of Physics University of Ioannina GR-45110 Ioannina Greece calissan@cc.uoi.gr

*Abstract-* We present the improved solar radio spectrograph of the University of Athens operating at the Thermopylae Satellite Telecommunication Station. Observations now cover the frequency range from 20 to 650 MHz. The spectrograph has a 7-meter moving parabola fed by a log-periodic antenna for 100 to 650 MHz and a stationary inverted V fat dipole antenna for the 20 to 100 MHz range. Two receivers are operating in parallel, one swept frequency for the whole range (10 spectrums/sec, 630 channels/spectrum) and one acousto-optical receiver for the range 270 to 450 MHz (100 spectrums/sec, 128 channels/spectrum). The data acquisition system consists of two PCs (equipped with 12 bit, 225 ksamples/sec ADC, one for each receiver). Sensitivity is about 3 SFU and 30 SFU in the 20-100 MHz and 100-650 MHz range respectively. The daily operation is fully automated: receiving universal time from a GPS, pointing the antenna to the sun, system calibration, starting and stopping the observations at preset times, data acquisition, and archiving on DVD. We can also control the whole system through modem or Internet. The instrument can be used either by itself or in conjunction with other instruments to study the onset and evolution of solar radio bursts and associated interplanetary phenomena.

## I. Introduction

Radio Spectrography of the solar corona, at decimeter, meter and decameter wavelengths, provides basic information on the origin and early evolution of many phenomena that later may reach the Earth. The Artemis IV solar radio spectrograph at Thermopylae Satellite Telecommunication Station (38.824166[0] N, 22.686666[0] E) of OTE (The Greek Telecommunication Organization) is a complete system [Maroulis D. et al, 1997], [Caroubalos C. et al, 2001], that received and recorded the dynamic spectrum of solar radio bursts on a daily basis [Caroubalos C. et al, 2004], from 100 to 650 MHz. These phenomena take place in lower corona. To follow them or to observe new phenomena in higher corona an extension to lower frequencies is needed, till the local ionosphere cut off that is about 10 MHz [http://iono.noa.gr]. Frequencies from 10-20 MHz have a lot of interference from strong short wave radio broadcastings and practically they are useless for the above purpose. At frequencies below 100 MHz, solar radio bursts are very strong [McLean D. et al, 1985 a] and can be observed easily, so a new simple but an efficient antenna was constructed for the 20-100 MHz band. Another reason to expand the low limit of the receiving frequencies is the collaboration with the STEREO/WAVES experiment [http://stp.gsfc.nasa.gov/missions/stereo/stereo.htm] because, i) its frequency range is from 10 kHz to 16 MHz, and ii) the FFR1 fixed frequency receiver works at 50 MHz.
Over the years, from the first installation of ARTEMIS IV (1996), some non-repaired hardware faults occurred which were removed by installing, modern digital systems, PCs and upgraded software. Finally, there are two antennas, two receivers and two PCs with A/D converters. (Fig. 1)

## II. The Two Antennas

The solar radio spectrograph has a parabolic and a dipole antenna (Figure 2).
   *The parabolic antenna* has a diameter of 7 m and is fed by a log periodic antenna for the 100-650 MHz band (High Band, HB). This antenna has a length of 2.25 m; its apex angle is 45[0] and consists of 13 dipoles. Input impedance is 50 Ω, almost independent of the frequency. Total gain $G_f$ (in dBi) depends on frequency $f$ (in MHz) and a second order polynomial curve fitting gives (Figure 3),

$$G_f = -4.10^{-5} f^2 + 0.043 f + 13.18 \qquad f=100, 101,\ldots,650 \qquad (1)$$

The antenna has a typical equatorial mounting and tracks the Sun on its daily and seasonal path.

After reception the signal goes through band pass filters (110-650 MHz), preamplifier (G=10 db, F=4 db) and multistage compensation amplifier (G=30 db) for the long transmission line. Every morning, the system starts automatically the antenna movement and performs self-calibration. The antenna is disconnected from the system and a white noise generator (constant $T=10^4$ K) is connected.

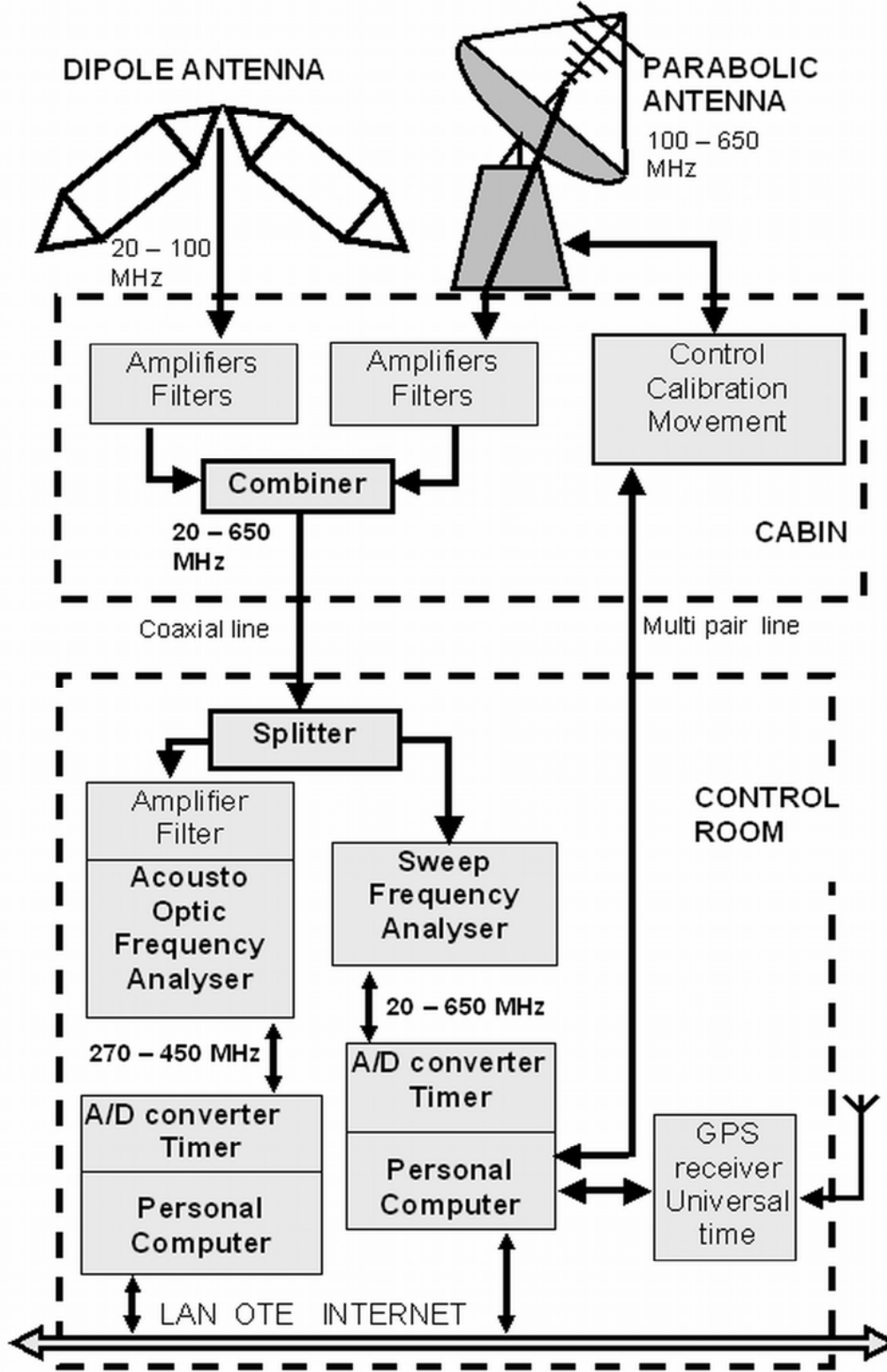

**Figure 1:** *ARTEMIS IV solar radio spectrograph block diagram.*

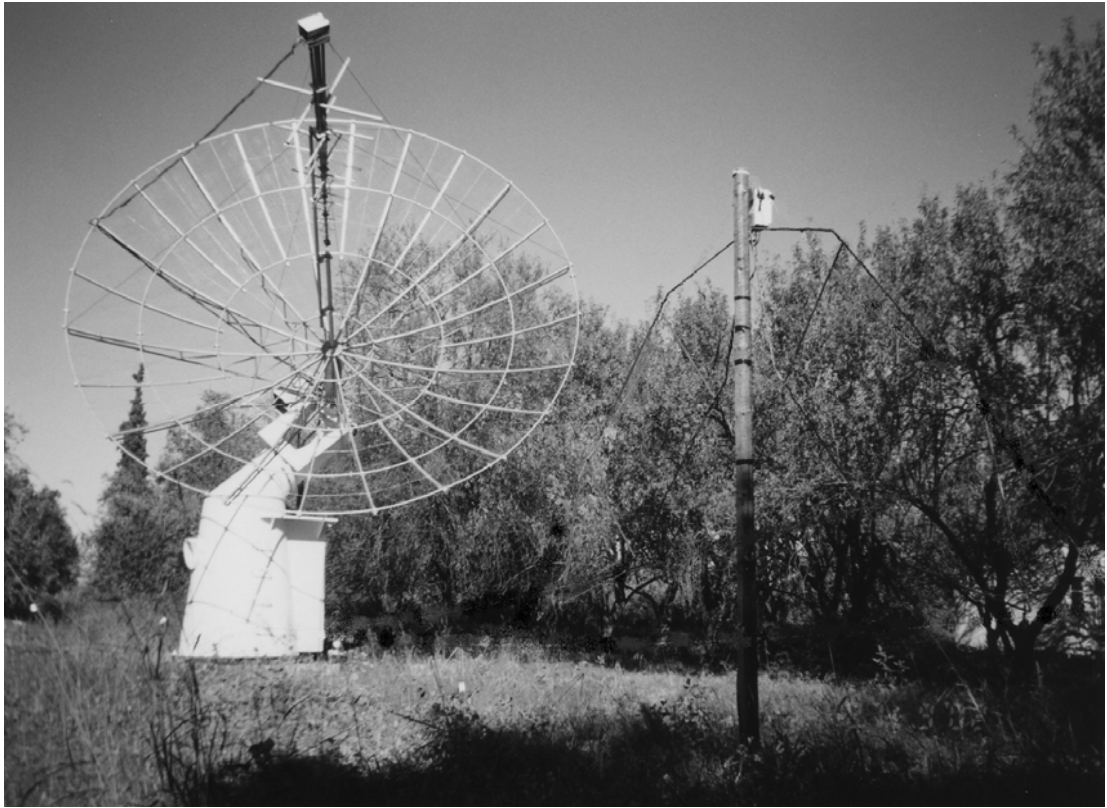

*Figure 2:* The two antennas at Thermopylae.

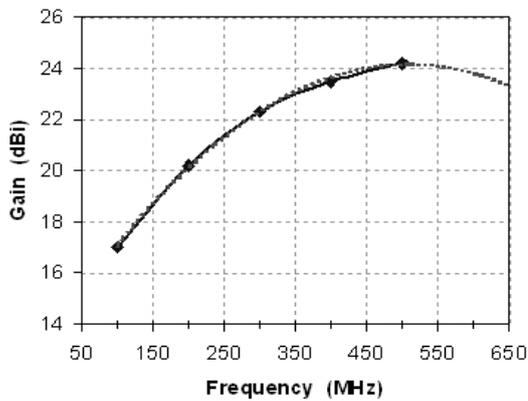

*Figure 3:* The gain of the HB antenna, square points are from measurements and dotted line after second order polynomial curve fitting

Afterwards the noise level is increased from $10^{1.5}$ to $10^8$ K, linearly with time, in 64 steps of 1 db. The duration of the calibration is about 1 min and the signal is recorded. This signal is used to understand the whole system behavior and to calibrate it. After the end of the calibration the antenna is connected again to the system.

*The dipole antenna* for the range 20-100 MHz (Low Band, LB) is a stationary very fat inverted V dipole, on the east-west vertical plane. Each leg has a length of 3.5 m and a width of 1 m. It is constructed by copper tube [Erickson W., 2003], [http://www.lofar.org]. The two legs form a $90^0$ angle and its apex is 3.6 m above ground. This antenna over a real ground ($\sigma$ =10 mS, $\varepsilon_r$ =14) was simulated by 4NEC2 software [http://www.si-list.org/swindex2.htm]. Standing wave ratio (SWR) for a 300 $\Omega$ transmission line is below 4 in the range 20-100 MHz (Figure 4). Total gain depends on frequency and horizontal coordinates ($\theta,\varphi$). The antenna gain towards a typical direction ($\theta,\varphi$)=($40^0,40^0$) is about 0 dBi (Figure 5). Figure 6 shows the antenna gain for all directions $(\theta, \varphi)$ for the frequency of 40 MHz. On the same diagram the daily sun path is indicated for three declination angles (winter $\delta=-23.45^0$, spring and autumn $\delta=0^0$, summer $\delta=23.45^0$), also the galactic center path is indicated. In Figure 7 we see that the antenna receives vertically polarized signals from low angles over the horizon from east to west and horizontally polarized signals from high angles over the horizon from south to north. So the antenna is almost omni directional and can receive the sun signals during all the day, but also a lot of interfering signals. After reception, the signal goes through a balanced band pass filter to reject strong interference from local FM broadcastings and strong short wave transmissions. The filter consists of a 11$^{th}$ order Chebychev low pass filter ($f_L$=90 MHz, pass band ripple 0.1 db) in series with a high pass ($f_H$=20 MHz) with optimized input and output sections to improve impedance matching between the antenna and the amplifier that follows. In Greece there are no TV broadcasts in the range 40-68 MHz, so this frequency range is free from strong interference. An active balun–broadband amplifier follows, composed of two CA 2832 [Motorola, 1991] hybrid wideband linear amplifiers operated in parallel; their outputs are combined with an RF transformer to drive the 50 Ω coaxial transmission line. With this design, we achieve 30 db signal amplification and impedance matching between the balanced high input impedance of the dipole antenna and the unbalanced 50 Ω transmission line. This entire configuration forms an active antenna for the 20-90 MHz band with input impedance of 50 Ω.

*A combiner* (Figure 1) combines the signals from the two antennas and drives them to the control room through a buried 70 m long, 50 Ω coaxial low loss transmission line.

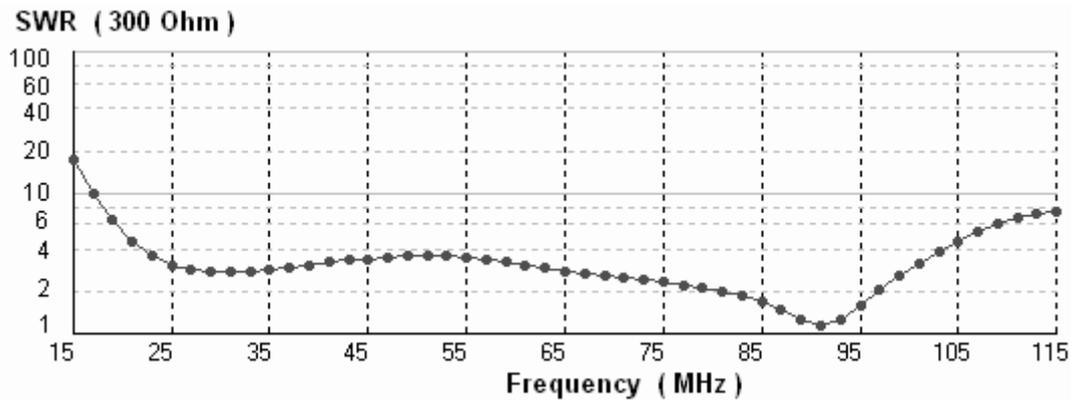

***Figure 4:*** *The voltage standing wave ratio of the LB antenna for a 300 Ω transmission line.*

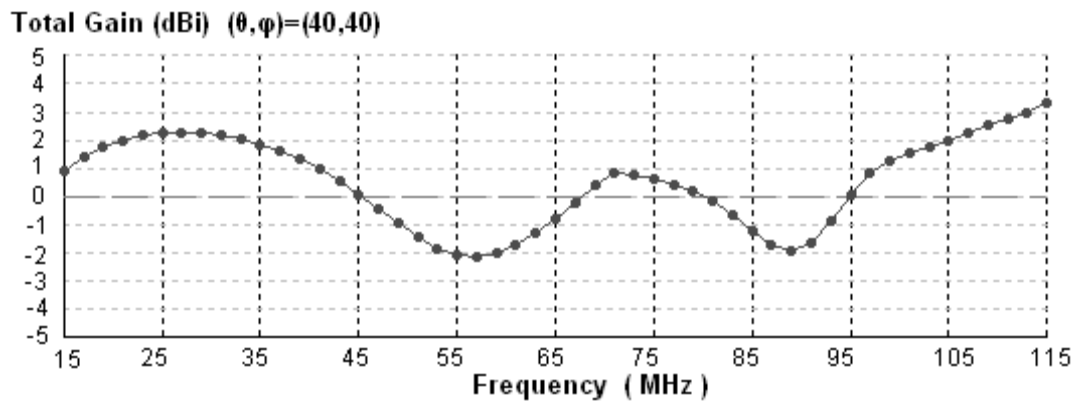

***Figure 5:*** *The total gain of the LB antenna towards an indicative direction $(\theta, \varphi)$ =$(40^0, 40^0)$.*

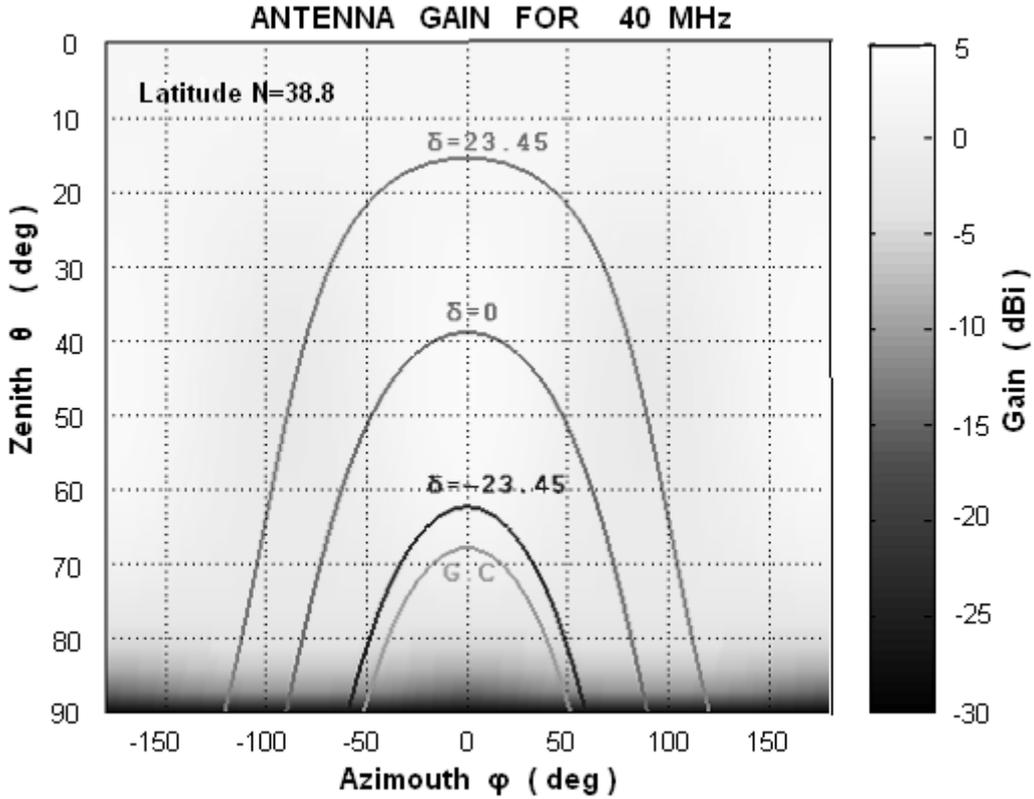

***Figure 6:*** *Total gain of the LB antenna for all horizontal directions (θ,φ) at 40 MHz. The sun and the galactic center path on the sky are also indicated for the Thermopylae site.*

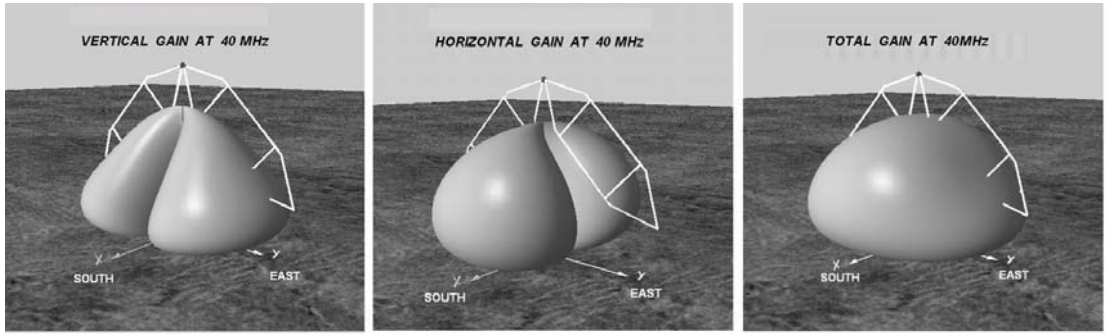

***Figure 7:*** *The structure of the LB antenna and its 3D far field polar diagram for the horizontal, vertical and total gain at 40 MHz. The antenna is almost omni directional.*

### III. The Two Receivers and Data Acquisition

The signal enters into the control room through the transmission line. A splitter splits the signal into two paths (Figure 1). *The first path* includes a modified commercial sweep frequency analyzer (Analyseur de Spectre Global or ASG) that covers the whole range from 20 to 650 MHz at 10 sweeps/sec with instantaneous bandwidth 1 MHz and dynamic range of 70 db.

The analogue output from the ASG (0-5 V dc) drives a 12-bit ADC card (Keithley KPCI 3100 ADC), [Keithley, 2002] on a PC (ASG PC). A pulse from a timer on the ADC card triggers every sweep at the ASG from low to high frequencies. The ADC takes 6300 samples/sweep (Figure 8). Every sample is an integer number between 0 and 4095. The samples are grouped in tens and the mean value is extracted for every group, so the whole spectrum (20 to 650 MHz) is divided into 630 channels with a resolution bandwidth of 1 MHz.

We also take care to avoid, in real time, strong interference from FM radio and TV broadcasting by rejecting the high data values at these a-priori constant noted frequencies. If in a channel (10 samples) some samples correspond to the constant frequencies of strong FM radio or TV broadcasting the averaging is done over the number of useful 'un-infected' measurements. If in one or more channels all the measurements are 'infected' their intensity is replaced by the linear interpolation of the two adjacent 'un-infected' channels. These arrangements lead to a high "signal from solar radio bursts to noise ratio".

The data from 5 successive sweeps (spectrums) form a block that is transferred to the hard disk with a universal time stamp. Simultaneously we have the dynamic spectrum on the PC screen. At the end of the day, or later, we can retrieve the daily activity and store it on a DVD. The daily data are about 0.5 Gbyte. Every day, before starting measurements, the ASG PC controls the calibration and the antenna movement through RS232 port; also this PC receives the Universal Time from a GPS and corrects its clock. This PC is equipped with a telephone line modem for telemetry purposes. Also there is an Ethernet connection for the other PC.

The software on the PC to take measurements, for the GPS time and for controlling the entire operation is in Delfi programming language, under Windows 2000 operating system. For each day there is a log text file with details for the various program operations.

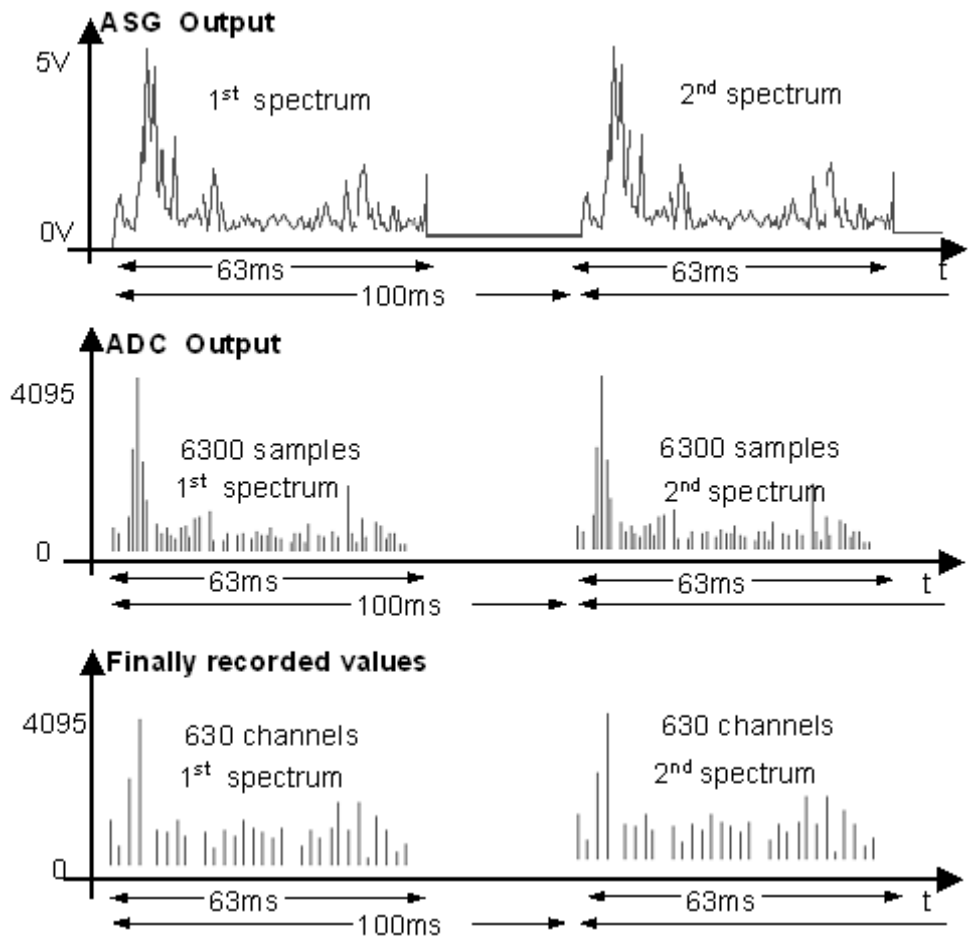

*Figure 8: From the ASG analogue output to the finally recorded values*

*The Second path* (Figure 1) includes a pass band filter (270-450 MHz), 30 db RF amplifier, and finally an acousto-optic frequency analyzer (Spectrograph Acousto-Optic, SAO) [Berg N. et al, 1996], [Rohlfs K., 1996] for 270-450 MHz with 25 db dynamic linear range, frequency resolution of 400 kHz and very fast frame rate of 100 Hz (one frame or spectrum in 10 msec). There is also a PC equipped with 12-bit ADC card (Keithley KPCI 3100). Every 10 msec the ADC takes 1024 samples from the SAO analogue output with a resolution of 12 bit in 0-5V dc range. Eight samples together form a group and the mean value is extracted for every group, so the range 270-450 MHz is divided into 128 channels with a resolution bandwidth of 1.4 MHz. This arrangement leads to a high "signal from solar radio bursts to noise ratio". The data from 50 successive sweeps (spectrums) form a block that is transferred to the hard disk with a universal time stamp. Simultaneously we have the dynamic spectrum on the PC screen. At the end of the day or later we can retrieve the daily activity and store it on a DVD. The daily data are about 1 Gbyte. The Universal Time is received every morning from the other PC through the Ethernet connection. The above frequency band of SAO has been selected because a lot of radio bursts with fine time structure are observed in this band and it is almost free from strong TV or FM broadcasting interference.

Figure 9 shows an example of ASG dynamic spectrum in grey scale (black for strong signals and white for weak). On the left we see the calibration signal from the noise generator, in the middle (inside the circle) a type III solar radio burst and on the right the greyscale in arbitrary units. Horizontal black or grey lines indicate strong FM radio and TV interference at constant frequencies.

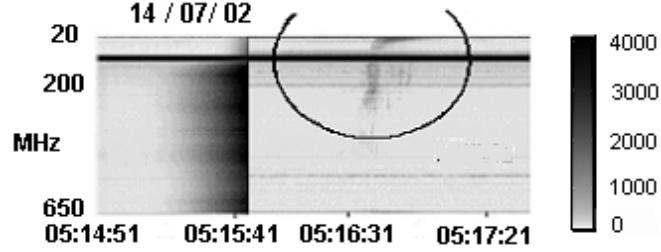

**Figure 9:** *Dynamic spectrum from the ASG. The calibration signal is shown at the left. The circle marks a type III solar radio burst. Black or grey horizontal lines are strong interference from FM radio and TV broadcastings at constant frequencies.*

### IV. Calibration

Calibration is performed to derive the relationship between the flux density (Wm$^{-2}$Hz$^{-1}$, or Solar Flux Unit, 1 SFU=10$^{-22}$ Wm$^{-2}$Hz$^{-1}$) that arrives from the solar radio burst at the antenna site, and all possible data values (0-4095) for 630 channels, from 20 to 650MHz with 1 MHz bandwidth. It consists of two steps, step one for the HB and step two for the LB.

*Step one:* a) The antenna temperature is derived from the recorded values of the calibration signal and the known noise temperature of the noise generator. b) The relation between the antenna temperature and the flux density is derived from the antenna temperature and the antenna far field pattern.

a) Before starting and after terminating the daily measurements, a calibration procedure takes place automatically. The parabolic antenna is disconnected from the preamplifier and a calibrated white noise generator is connected. The noise temperature of the calibration signal increases automatically linearly with time from 10$^{1.5}$ to 10$^{8}$ K in 64 db steps. Every step lasts for 0.8sec. Let $T_{f,s}$ be the noise temperature for every channel (frequency) $f$ ($f$=100, 101, 102, ...,650) and for every temperature step $s$ ($s$=1,2,...,64).

$$Log(T_{f,s}) = 1.5 + s\frac{7.5}{64} \qquad (2)$$

At the same time the output $m_{i,s}$ from the analyser is recorded by the ASG PC in integer values, ($m_{i,s}$ from 0 to 4095) for every channel $f$ and for every temperature step $s$. In every step there are eight values $m_{i,s}$ which are averaged, to minimize noise, and give the mean value $M_{i,s}$

So for every value of $M_{i,s}$ there is a corresponding antenna temperature $T_{i,s}$. After interpolation we derive the relation

$$T_{f,j}=\tau_{HB}\ (f,j) \qquad \text{for } f\text{=100, 101,102,...,650 and } j\text{=0,1,2, ,4095} \qquad (3)$$

b) From the antenna temperature $T_{f,j}$ we can calculate the flux density $S_{f,j}$ using the relation [Kraus, 1988]:

$$S_{f,j} = \frac{2\,k\,T_{f,j}}{A_f} \qquad (4)$$

where: $k$ the Boltzmann constant and $A_f$ the effective aperture of the antenna at frequency $f$. The effective aperture $A_f$ of the antenna is [Kraus, 1988]:

$$A_f = \frac{G_f\lambda_f^{\ 2}}{4\pi} \qquad (5)$$

where: $G_f$ the gain of the main beam of the HB antenna and $\lambda_f$ the wavelength at the frequency $f$ ($\lambda_f = 3 \times 10^8/f$).

The antenna gain is known from (1) and is independent from the Sun position because the antenna every time points to the Sun. So from relations (1), (5), (3) and (4) we derive the relation:

$$S_{f,j} = \sigma_{HB}(f,j) \qquad \text{for } f=100, 101,102,\ldots,650 \text{ and } j=0,1,2, \ldots,4095 \qquad (6)$$

Figure 10 shows this relation for three indicative frequencies $f$. Nonlinear segments at the beginning and at the end of the curves are owing to the nonlinear effects of the amplifiers.

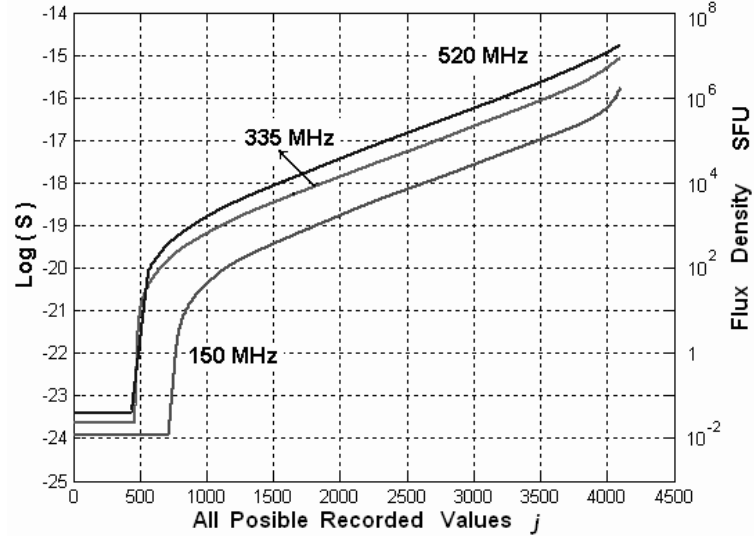

***Figure 10:*** *Calibration curves for three indicative frequencies for the HB antenna.*

*Step two*: a) The antenna temperature is derived from the recorded values of the calibration signal and the known noise temperature of the noise generator. b) The relation between the antenna temperature and the flux density is derived from the antenna temperature and the antenna gain towards the direction of the Sun.

a) The previous procedure (disconnection of the antenna from the filter-amplifier, connection of the calibrated white noise generation) has been applied at once when the dipole antenna was installed and we derive the relation:

$$T_{f,j} = \tau_{LB}(f,j) \qquad \text{for } f=20, 21,\ldots,100 \text{ and } j=0,1,2, \ldots,4095 \qquad (7)$$

b) The direction $(\theta,\varphi)$ of the Sun is calculated from the Universal Time at which a radio event has been recorded [Kinelson et al, 1995]. For this direction, the antenna gain $G_f$ is calculated by simulation for every frequency $f$ ($f=20, 21,\ldots,100$), as in figure 5. The flux density $S_{f,j}$ is derived using this value of $G_f$ and relations (5) and (4). Finally we have the relation

$$S_{f,j} = \sigma_{LB}(f,j) \qquad \text{for } f=20, 21,\ldots,100 \text{ and } j=0,1,2, \ldots,4095 \qquad (8)$$

The results show that the sensitivity is about 3SFU in LB and about 30SFU in HB with dynamic range of 45db. We can thus detect only solar radio bursts and no radio signal from the quiet Sun that produces lower flux density at these frequencies.

Detecting the noise, for a whole day, from the Galactic Center in LB, and finding values that are referred to in the bibliography [Cane H., 1979], the method of calibration and the results from the simulation of the LB antenna have been verified.

### V. Data analysis

Data file for a solar radio event is a MxN array where M is the number of channels (M=630 for ASG and M=128 for SAO) and N the number of spectra that occupy the event (for ASG we have one spectrum for every tenth of the sec. and for SAO we have one spectrum for every hundredth of the sec.). The elements of the array are integer numbers (0-4095) that represent the intensity of the event.

Data processing involves: a) filtering the signal to remove interference, b) finding the flux density of the dynamic spectrum and c) calculating the power density (Wm$^{-2}$) for the whole range from 20 to 650 MHz.

Man made radio interference is increasing every year as social activities are increasing and there is no respect shown to frequencies allocated to radio astronomy; all these bring on difficulties for observing the solar radio bursts, clearly. There are various types of man made and physical interference, such including:

1. Narrowband 'instant' interference, its bandwidth being narrower than the channel bandwidth (1 MHz) due to narrow band FM or AM mobile communications at various frequencies; it occupies only one channel with maximum duration of some minutes. They appear as small narrow horizontal lines.

2. Narrowband 'lasting' interference, its bandwidth being narrower than the channel bandwidth (1 MHz) due to FM or AM radio broadcasting at fixed frequencies; it occupies only one channel, but is present all the time or for some hours. They appear as very long narrow horizontal lines.

3. Wideband 'lasting' interference, its bandwidth being larger than the channel bandwidth (1 MHz) due to TV radio broadcasting at fixed frequencies; it occupies some channels (4-10) and is present all the time or for some hours. Also, this kind of interference appears when there is narrowband lasting interference at adjacent channels as in the FM radio broadcasting band (88-108MHz) or TV broadcasting bands. They appear as very long wide horizontal lines

4. Ultra wideband instant interference that occupies all channels from 20MHz to some hundred MHz due to lightning with maximum duration of about one second. They appear as narrow vertical lines.

5. Galactic background broadband radiation dominates the system noise at the 20-100MHz band. It is stronger at the lower part of the band and dominates when the galactic centre is on the sky during daytime from November to February [McLean et al, 1985 a].

Contrary to all these, solar radio events of various types (I, II, III, IV and V) are broadband with a duration of some seconds to hours [McLean D. et al, 1985 b]. Solar radio events can be recovered from the data, keeping in mind the above factors and by using standard image processing with 2D median, low pass or high pass filtering. A more efficient filtering is achieved by using, for every spectrum, moving median filters with variable bandwidth. Figure 11 shows an example.

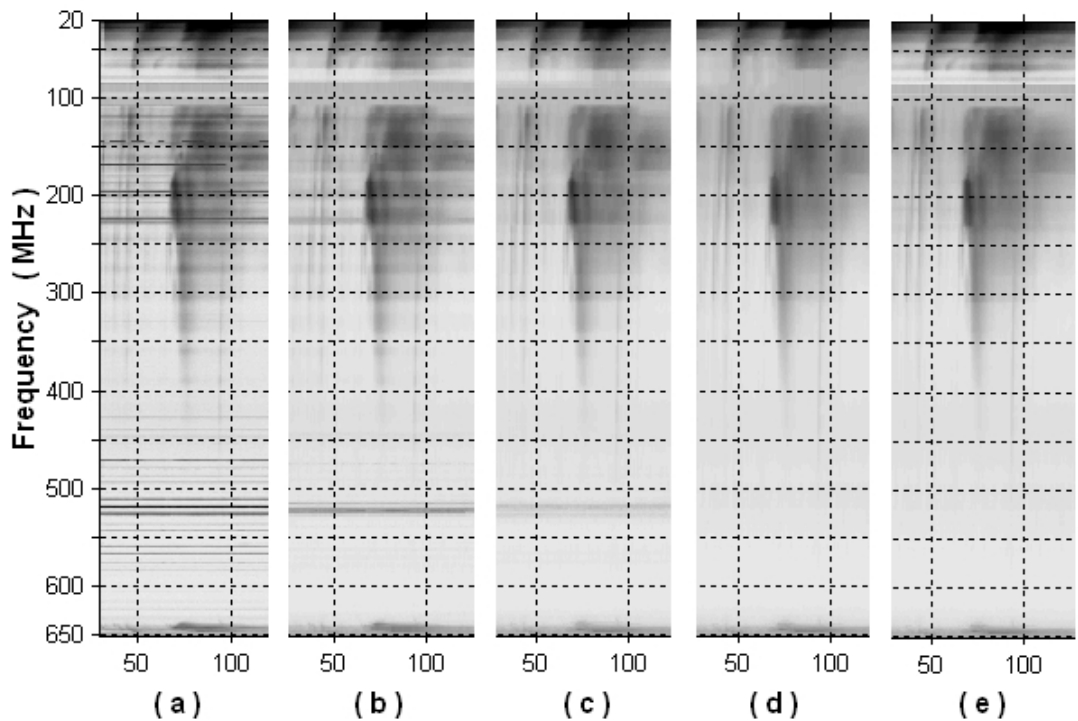

*Figure 11: The influence of the moving median filtering on dynamic spectrum. (a) Initial dynamic spectrum. (b) Moving median filtering with constant bandwidth 5. (c) Moving median filtering with constant bandwidth 15. (d) Moving median filtering with constant bandwidth 25. (e) Moving median filtering with variable bandwidth. Large bandwidth rejects all the interference and a lot of details in the solar radio burst. Small bandwidth allows details in the solar radio burst but cannot reject interference. Using variable bandwidth the spectrum is divided into regions and we apply moving median filtering with variable bandwidth as follows: region 20 − 80 MHz, 80 − 400 MHz and 400 − 650 MHz with bandwidth 2, 17 and 25 respectively.*

After filtering, we use the calibration relations (6) and (8) to derive the dynamic spectrum flux density in SFU. Integrating these values for every spectrum we calculate the total power density in Wm$^{-2}$ for the whole range 20-650MHz. An example is shown in Figure 12.

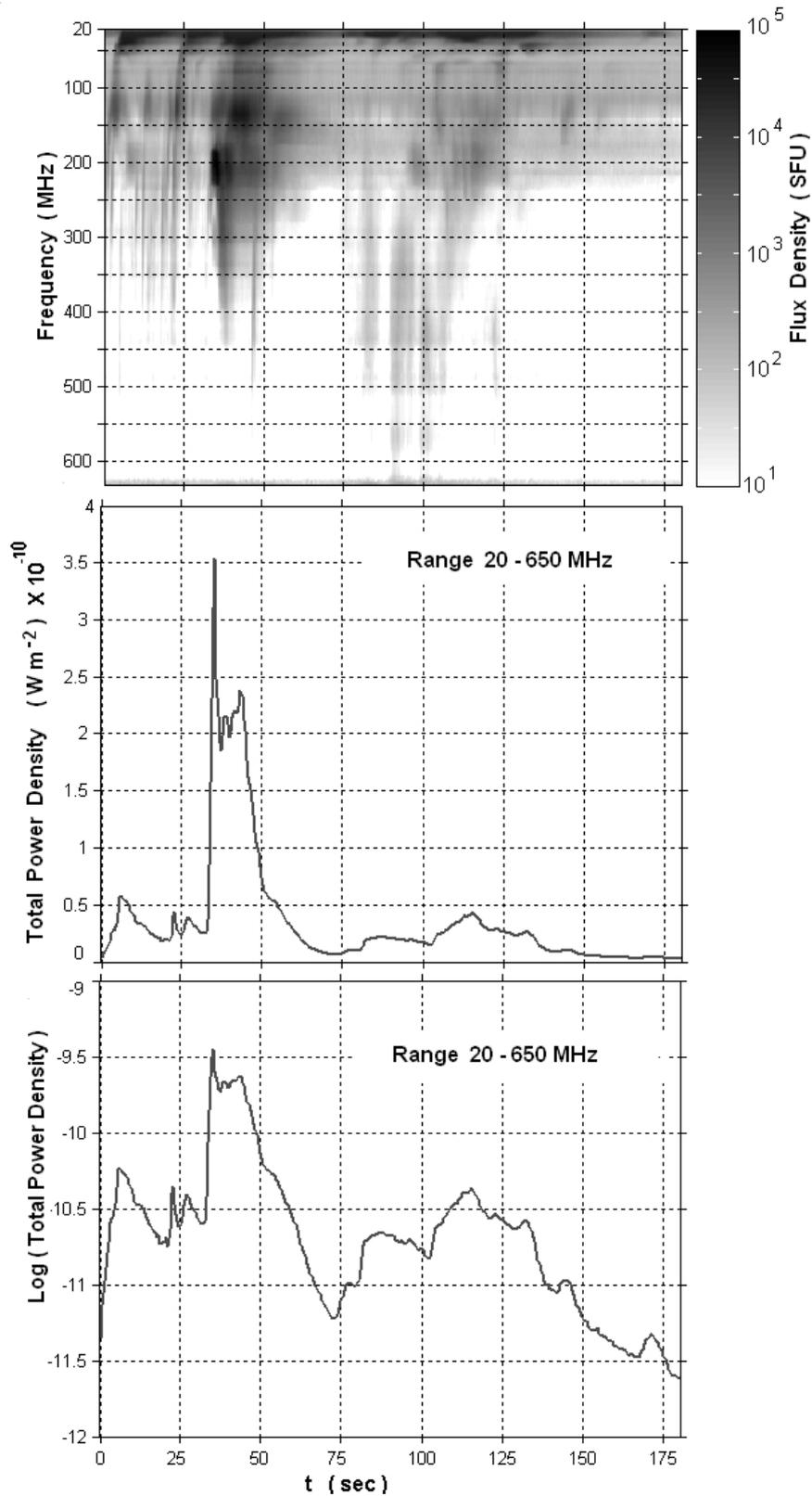

**Figure 12:** *Dynamic spectrum of type III and V solar radio burst from 08:05:00 UT on 26 April 2003. The graphs show the Total Power Density and its logarithm for the range 20-650MHz.*

This instrument has a very high time and frequency resolution from decimeter to decameter wavelengths with a very good signal to noise ratio that permits fine structure detection [Tsitsipis P. et al, 2001].

Figure 13 shows the dynamic spectra of 10min. long solar radio event on 28/10/2004 recorded by ARTEMIS IV ASG solar radio spectrograph. Grey scale on the right shows the signal intensity in arbitrary units. Figure 14 is a 2min part of the above event. It shows the differential dynamic spectrum derived after special 2D filtering where we can distinguish fine structure, which indicates ascending and descending electron beams in the solar corona. We observe a type II solar radio burst with harmonic bands and herringbone structure. Figure 15 and figure 16 show the same event as recorded by SAO in the 270-450MHz band where more details are seen in the frequency domain.

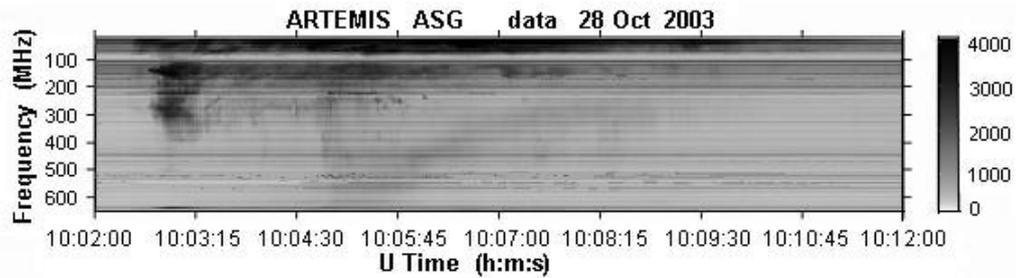

***Figure 13:*** *Dynamic spectrum of a large solar event recorded by ASG from 10:02:00 to 10:12:00 on 28 Oct. 2003. Grey scale on the right is in arbitrary units*

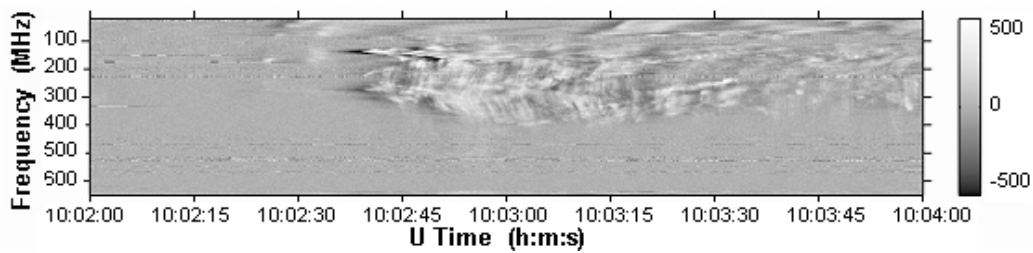

***Figure 14:*** *Detail of the figure 13, shows the time differential dynamic spectrum from 10:02:00 to 10:04:00, derived after special 2D filtering. Ascending and descending lines are produced by ascending and descending electron beams in the solar corona.*

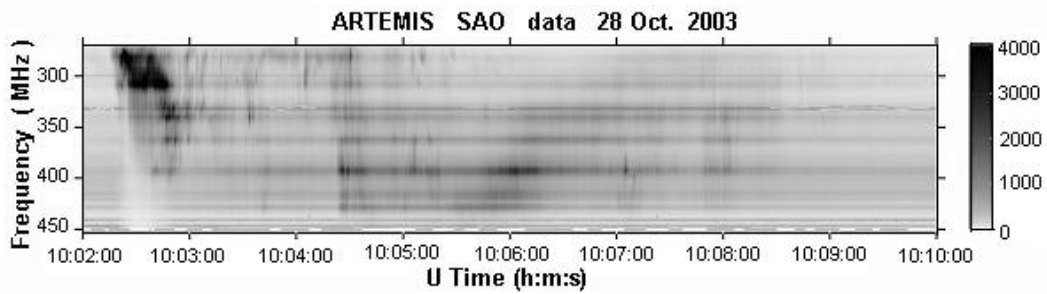

***Figure 15:*** *Dynamic spectrum of the same large solar event recorded by SAO from 10:02:00 to 10:10:00 on 28 Oct. 2003. Grey scale on the right is in arbitrary units*

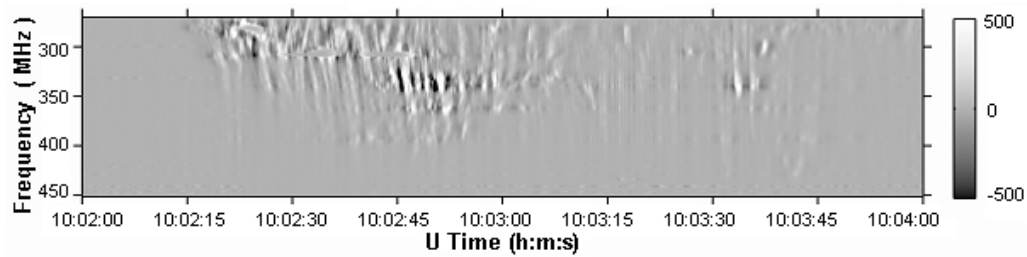

***Figure 16:*** *Detail of the figure 15, shows the time differential dynamic spectrum from 10:02:00 to 10:04:00, derived after special 2D filtering.*

## VI. Conclusions and perspectives

We have designed and constructed a system, with two antennas and a frequency sweep radio spectrograph, that observes solar radio bursts in the range 20-650 MHz with a time resolution 100 msec, frequency resolution of 1 MHz, sensitivity of 3 SFU and 30 SFU in the 20-100 MHz and 100-650 MHz range, respectively. Using an acousto-optic radio spectrograph in the range 270-450 MHz, time resolution of 10 msec. and frequency resolution of 1.4 MHz has been achieved.

Future perspectives are the use of the instrument for routine observations of solar radio bursts, the construction of new dipole antennas perpendicular to the present so that we can make polarimetric measurements, as well as the extension of the SAO frequency range.

## VII. Acknowledgements

This work was supported in part by the program PLATON N° 05341PJ for the French-Greek collaboration and the program EPEAEK of the Greek Ministry of Education and Religious Affairs and the European Union in the framework of the research project Archimedes II: "Receiving and Processing the Signal from Solar Radio Bursts". We also thank the Greek Telecommunication Organization (OTE) for hosting the equipment of ARTEMIS IV in the area of the earth satellite station "Thermopylae".